\documentclass[aps,prl,twocolumn,showpacs,superscriptaddress,groupedaddress]{revtex4}
\usepackage{amsmath, amsthm, amssymb}
\usepackage{latexsym}
\usepackage{amsfonts}
\usepackage{graphicx}
\usepackage{amsbsy}
\usepackage{float}
\usepackage{amsmath}

\begin{document}

\title{Network Risk and Forecasting Power in Phase-Flipping Dynamical
  Networks}

\author{B.~Podobnik}
\affiliation{Center for Polymer Studies and Department of Physics, Boston
University, Boston, MA 02215, USA}

\affiliation{Faculty of Civil Engineering, 
University of Rijeka, 51000 Rijeka,  Croatia}

\affiliation{Zagreb School of Economics and Management, 10000 Zagreb, 
 Croatia}

% \affiliation{Faculty of Economics, 
%University of Ljubljana, 1000 Ljubljana,  Slovenia}

\author{A.~Majdandzic}
\affiliation{Center for Polymer Studies and Department of Physics, Boston
University, Boston, MA 02215, USA}

\author{C.~Curme}
\affiliation{Center for Polymer Studies and Department of Physics, Boston
University, Boston, MA 02215, USA}

\author{Z. Qiao}
\affiliation{NUS Graduate School for Integrative Sciences and Engineering,
 NUS, Singapore 117456, Singapore.}
\affiliation{Department of Physics and Center for Computational Science and 
Engineering,
NUS, Singapore 117456, Singapore}

\author{W.-X. Zhou} 
\affiliation{School of Business, School of Science, and Research Center for
Econophysics, East China University of Science and Technology, Shanghai 200237,
China}

\author{H.~E.~Stanley}
\affiliation{Center for Polymer Studies and Department of Physics, Boston
University, Boston, MA 02215, USA}

\author{B.~Li}
\affiliation{NUS Graduate School for Integrative Sciences and Engineering,
 NUS, Singapore 117456, Singapore.}
\affiliation{Department of Physics and Center for Computational Science and 
Engineering,
NUS, Singapore 117456, Singapore}

\affiliation{
%Center for Phononics and Thermal Energy Science, 
School of Physics Science and Engineering, Tongji University, Shanghai,
200092, P R China}

\begin{abstract} 

{\bf In order to model volatile real-world network behavior, we 
  analyze  phase-flipping dynamical 
  scale-free network in which nodes and links fail and recover.  We
  investigate how stochasticity in a parameter governing the recovery
  process affects phase-flipping dynamics, and find the probability
  that no more than $q\%$ of nodes and links fail. We derive higher
  moments of the fractions of active nodes and active links, $f_n(t)$
  and $f_{\ell}(t)$, and define two estimators to quantify the level of risk
  in a network.  We find hysteresis in the correlations of $f_n(t)$ due
  to failures at the node level, and derive conditional probabilities
  for phase-flipping in networks.  We apply our model to economic and
  traffic networks.}

\end{abstract}


\maketitle 

Across a broad range of human activities---from medicine, weather, and
traffic management to intelligence services and military
operations---forecasting theories help us estimate the probability of
future outcomes. In general, the greater the uncertainty of outcome, the
more crucial that we be able to forcast future behavior.  Since the
nodes of many dynamic
systems~\cite{Watts,Barabasi99,Adamic,Albert00,Cohen00,Holme12,Pierra,Krapivski,Milo,Garlaschelli03,AlbertRMP,Buldyrev10,GaoNP12,Dorogotsev,Newman,Dorogovtsev00,Helbing,Vespignani01,Vespignani12,BP12,ParshaniPNAS},
such as traffic patterns and physiological networks, periodically fail
and then recover---a disease spreads through an organism and then after
a finite period of time the organism recovers---the forecasting power~\cite{Vespignani05,Vespignani09,Kolar} is highly
relevant---it allows us to estimate the probability of future node and
link failure and recovery and to quantify the level of risk in any given
dynamic network.

A recent paper details how the nodes in dynamic regular networks and
Erd\H{o}s Renyi networks (i) inherently fail, (ii) contiguously fail due
to the failure of neighboring nodes, and (iii) recover \cite{Antonio03}.
These networks exhibit phase-flipping between ``active'' and
``inactive'' collective network modes.  Here we analyze networks
 with highly heterogeneous degree distributions   and we describe
scale-free phase-flipping networks in which nodes and links fail and
recover. We describe the collective behavior of these networks using two
time-dependent network variables: the fraction of active nodes $f_n(t)$
and the fraction of active links $f_{\ell}(t)$.  We 
 place an emphasis on forecasting in dynamic networks---we want to 
  calculate  how many nodes will fail at any
future time $t$---and quantify how risky networks are. \\  
%\begin{itemize}
%\item[{(i)}] 
 (i) At each time $t$ any node in the system can independently
  fail, breaking its links with all other nodes, with a probability $p$.
  The internal failure state of node $i$ we denote by spin
  $|s_i\rangle$ (if  $i$ is active, $|s_i\rangle = | 1 \rangle$).  \\ 
%\item[{(ii)}] 
 (ii) The external failure states we denote by spin
  $|S_i\rangle$, where $|S_i\rangle$ is $|1\rangle$ if node $i$ has more
  than $T_h\%$ active neighbors, and $|0\rangle$ (for a subsequent time
  $\tau' = 1$) with probability $p_2$ if $\leq T_h~\%$ of $i$'s
  neighbors are active.  For scale-free networks, a percentage threshold
  $T_h$ is the more appropriate choice than the constant $T_h$ used in
  Ref.~\cite{Watts,Antonio03}. Node $i$---described by the two-spin
  state $| s_i,S_i \rangle$---is active only if both spins are up (1),
  i.e., if $| s_i,S_i \rangle = | 1, 1 \rangle$. \\ 
%\item[{(iii)}]
 (iii)   After a time period $\tau$, the nodes recover from
  internal failure. Usually $\tau$ is random, but we also analyze the
  case when $\tau$ is constant~\cite{Antonio03}. \\ 

%\end{itemize}
 
\begin{figure}[b]
\centering \includegraphics[width=0.23\textwidth]{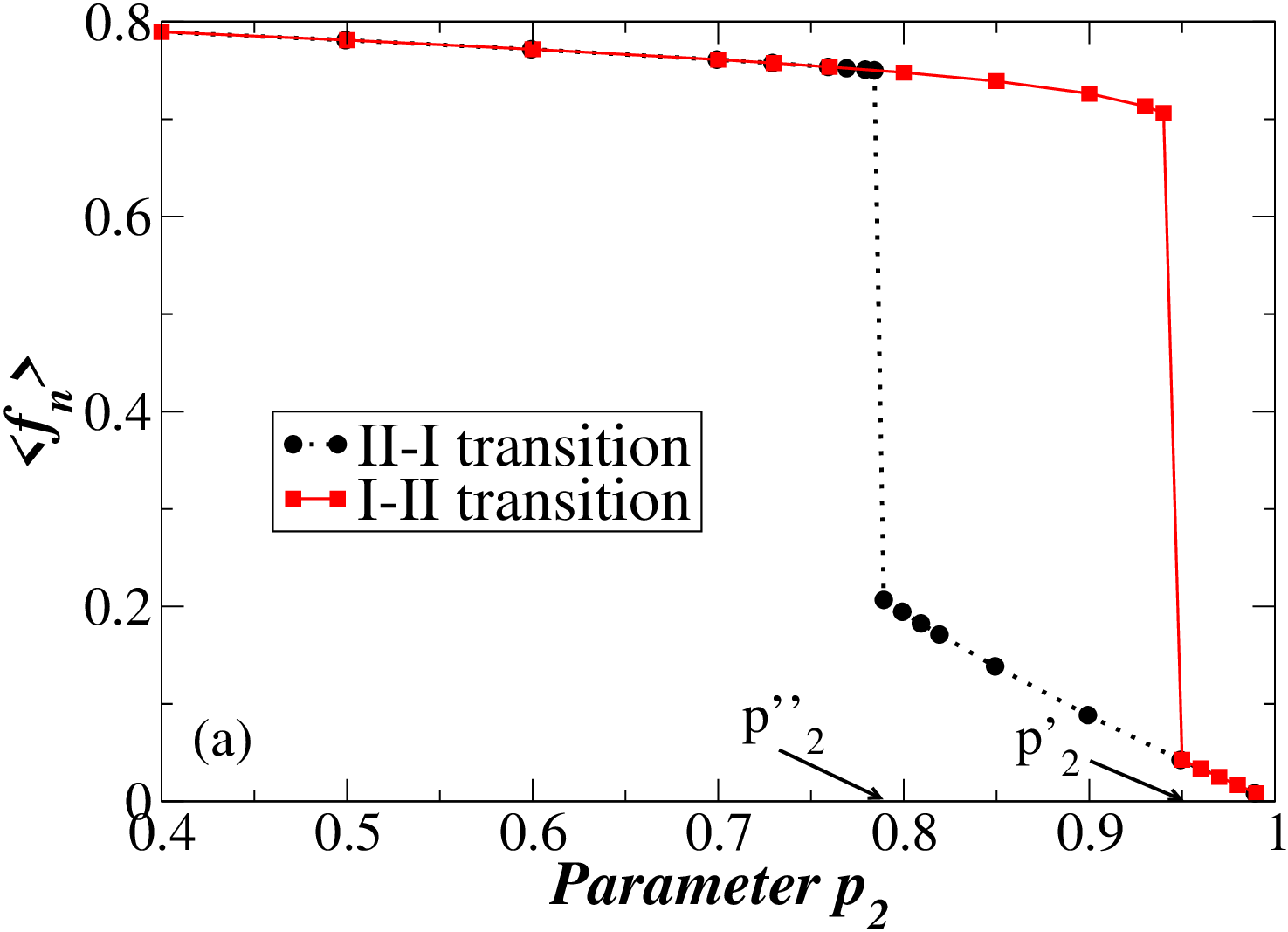}
\centering \includegraphics[width=0.23\textwidth]{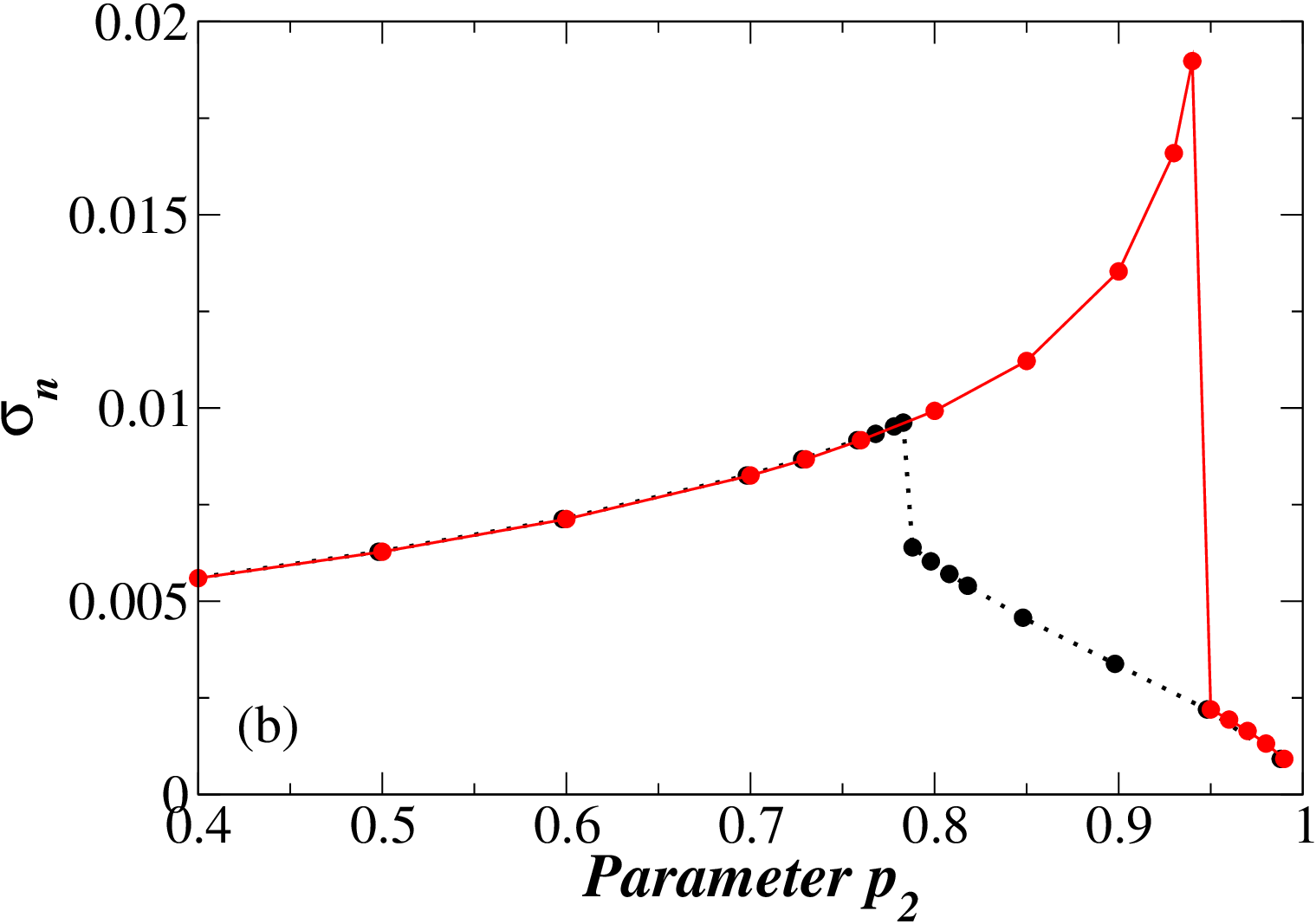}\\
\centering \includegraphics[width=0.23\textwidth]{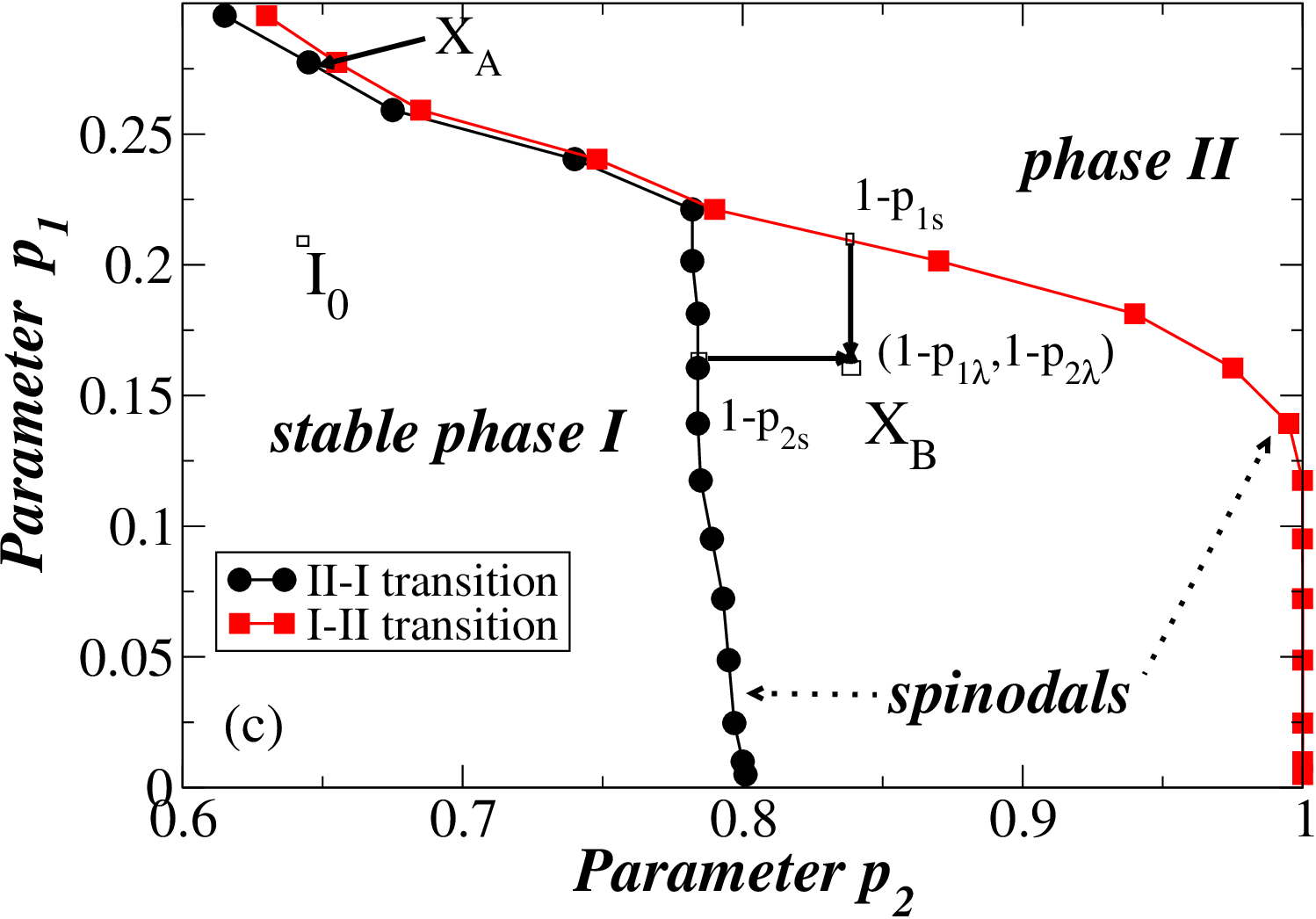}
\centering \includegraphics[width=0.23\textwidth]{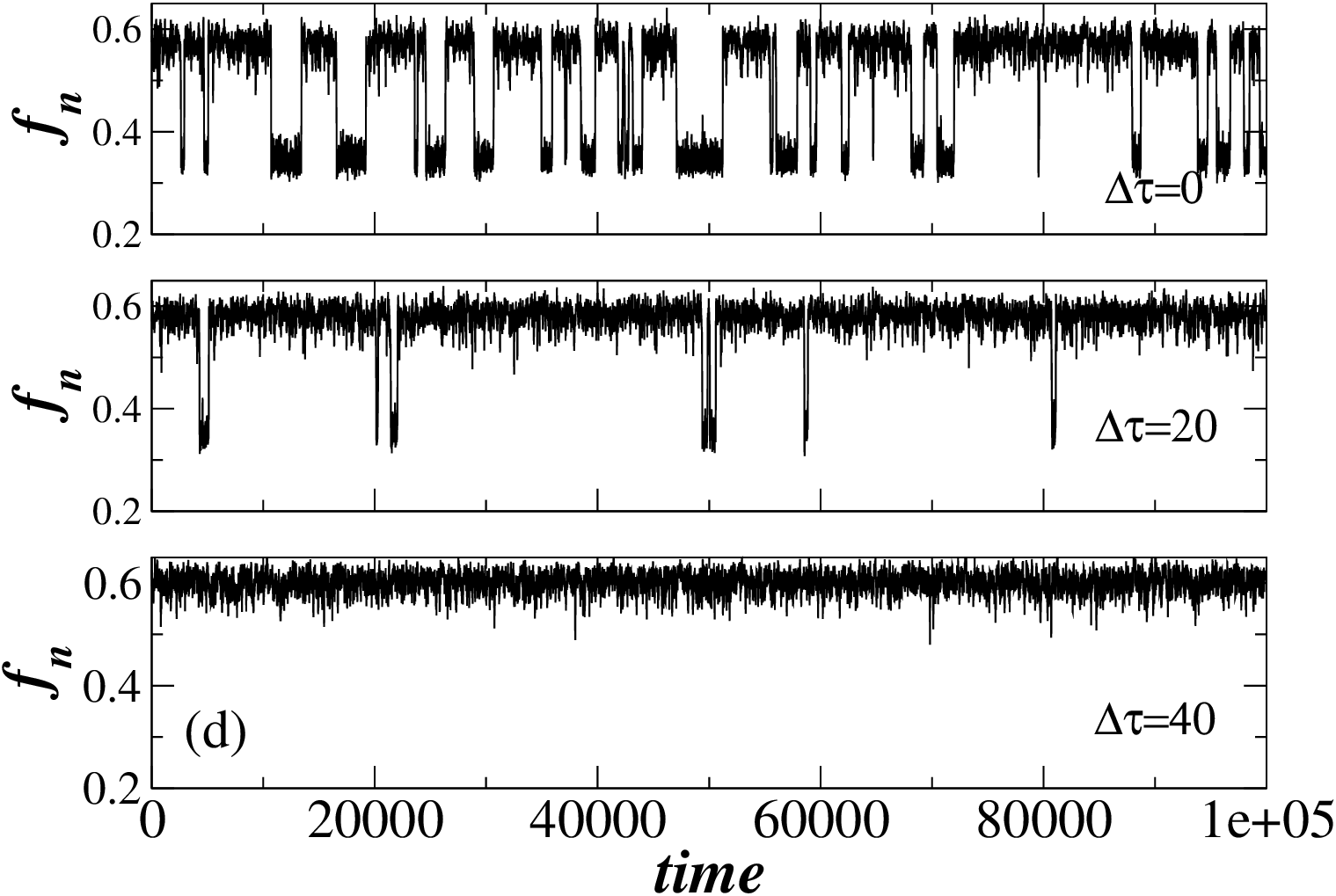}
\caption{Network statistics used in analyzing (in)stability 
  in the (i)--(iii) networks.  In (a)--(c) we start with a BA network
  with $N=10,000$ and $\langle k \rangle = 3$, and then introduce the
  (i)--(iii) network, where $T_h = 50\%$ and $\tau = 50$.  Fixing $p_1$
  and increasing $p_2$, for each ($p_1,p_2$) we calculate the fraction
  of active nodes, $f_n(t)$.  We show hystereses for two network
  statistics: (a) the average $f_n(t)$, $\langle f_n \rangle$, and (b)
  the standard deviation of $f_n(t)$, $\sigma_n$.  We
  use $p_1=0.2~(p=0.004)$ in two directions: from $p_2=0$ to
  $p_2=1$, and then from $p_2=1$ to $p_2=0$. As $p_2 \rightarrow p'_2$,
  $\langle f_n \rangle$ and $\sigma_n$ exhibit first-order transitions.
  (c) Emergence of hysteresis in ($p_1,p_2$) space.  The hysteresis
  point $X_A$ is characterized by $\tau = 50$, $p = 0.0065$ ($p_1 =
  0.277$), and $p_2 = 0.65$. (d) If $\tau$ is a random variable 
  from a homogeneous pdf, $H(\tau_0 - \Delta \tau, \tau_0 + \Delta \tau)$,
  with $\tau_0 = 50$, the phase-flipping phenomenon gradually dissappears
  with increasing $ \Delta \tau$.  }
\label{1}
\end{figure}

Estimating how far the parameters of a dynamic system are from the area
in parameter space characterized by high instability is crucial.  For
the network described in (i)--(iii) above, we arbitrarily choose
parameters $p_1$ (related to $p$ by $p_1=1-\exp(-p\tau)$
\cite{Antonio03}) and $T_h$. We then destabilize the network by
increasing $p_2$, causing it to transition from phase I with
predominantly active nodes to phase II with predominantly inactive
nodes.  In Fig.~\ref{1}(a), for varying $p_2$, the first network
statistic---the average $f_n(t)$, $\langle f_n \rangle$---gradually
decreases for $p_2 \in (0, p'_2)$ and then, at $p_2 \approx p'_2$,
$\langle f_n \rangle$ shows a sudden network crash---a first-order phase
transition.  In Fig.~\ref{1}(b) for $p_2 \in (0, p'_2)$ the second
network statistic---the standard deviation of $f_n(t)$,
$\sigma_{n}$---becomes increasingly volatile.  During network recovery,
in Figs.~\ref{1}(a) and \ref{1}(b) both $\langle f_n \rangle$ and
$\sigma_n$ follow a first-order phase transition, but at a value
$p_2=p''_2$, which differs from $(p_2=p'_2)$ obtained during the I--II
transition.  Because $\langle f_n \rangle$ and $\sigma_n$ are dependent
upon the initial node spins in the network, $p'_2 \ne p''_2$ implies the
existence of hysteresis \cite{NSR12,Angeli,Blanchard,Das}.  To estimate
the part of the $(p_1,p_2)$ phase space that is unstable, we calculate
the discontinuity $(p'_2, p''_2)$ values for varying $p_1$ values [see
  Fig.~\ref{1}(b)].  Figure~\ref{1}(c) shows a hysteresis with two
discontinuity lines (spinodals) in the ($p_1,p_2$) space.  The closer
the ($p_1,p_2$) of a network is to the left spinodal in Fig.~\ref{1}(c),
the less stable the network.
  
Reference~\cite{Antonio03} reports that introducing both a dynamic
recovery with a (constant) parameter ($\tau \ne 0$) and a stochastic
contiguous spreading ($p_2 \ne 1$) leads to spontaneous collective
network phase-flipping phenomena.  Figure~\ref{1}(d) shows the fraction
of active nodes $f_n(t)$ for constant $\tau$ ($\Delta \tau = 0$) 
that corresponds to the
volatile state $X_A$ shown in Fig.~\ref{1}(c).  Figure~\ref{1}(d) shows
that if $\tau$ is not constant but a random variable from a homogeneous probability
distribution function (pdf), the phase-flipping
phenomenon and thus the collective network mode disappears with
increasing $\Delta\tau$ (increasing stochasticity in $\tau$).  Beginning
with the relation $p_1 = 1 - \exp(-p\tau)$ \cite{Antonio03} when $\tau$
is constant, we confirm this result.  Suppose a network is initially set
at a phase-flipping state $X_A$ [Fig.~\ref{1}(c)].  If $\tau$ follows a
homogeneous pdf, $H(\tau_0 -\Delta \tau, \tau_0 +\Delta \tau)$, we
easily derive the average parameter $p^*_1 \equiv p_1(\Delta \tau)$ as
\begin{equation} 
E(p^*_1) = 1 - \exp(-p \tau_0) \sinh(p \Delta \tau)/p\Delta\tau,
\end{equation} 
 and
the average deviation of $p^*_1$ from $p_1 \equiv p_1(\Delta \tau=0)$,
$E(p^*_1) - p_1 = \exp(-p \tau_0) [ 1 - \sinh(p \Delta
  \tau)/p\Delta\tau] < 0$.  With increasing $\Delta \tau$, $E(p^*_1) -
p_1$ decreases and $E(p^*_1)$ moves from a volatile network regime
($X_A$) to a more stable network regime. Hence at $X_A$ the less
dispersed $\tau$ is (and also $p_1$), the more pronounced the
phase-flipping. Hereafter, we analyze networks with constant $\tau$. 
 
We next explore the diagnostic and forecasting power of dynamic
networks.  When internal ($X$) and external ($Y$) failures are
independent, according to probability theory $P(X \cup Y) =P(X) + P(Y) -
P(X) P(Y)$, from which Ref.~\cite{Antonio03} calculates the probability
$a = a(p,p_2,T_h) \equiv P(X \cup Y)$ that a randomly chosen node $i$ is
inactive
\begin{equation} 
 a = p + p_2 (1 -p) \Sigma_{k}P(k) E(k,m,a), 
 \label{a}
\end{equation} 
equal to the fraction of inactive nodes, $a = 1 - \langle f_n
\rangle$. Clearly,  the internal and external failures  are 
 only approximately independent~\cite{Antonio03}. 
Here $P(k)$ is the degree distribution, $T_h$, $p$, and $p_2$
are described in (i)--(ii) above, $m\equiv T_h k$, $E(k,m,a) \equiv
\Sigma_{j=0}^{m} a^{k-j} (1 - a)^{j} {k \choose k - j}$ is the
probability that the neighborhood of node $i$ is critically damaged.
% and $k$ is the number of links still active for node $i$. 
 For a network
with $N$ nodes, each with probability $(1-a)$ of being active, using a
binomial distribution we obtain any moment of $f_n$ of order $q$,
\begin{equation} 
 \langle  f^q_n \rangle 
  \equiv \Sigma_{j=0}^N (\frac{j}{N})^q a^{N-j} (1-a)^j  {N \choose j}, 
 \label{moment}
\end{equation} 
that is, for large values of $N$, $\langle f^q_n \rangle \approx \int dx
x^q G(x,\mu=1-a,\sigma=\sqrt{a(1-a)/N})$---G stands for Gaussian.  The
dependence of $f^q_n$ on $a$ explains why both $\langle f_n \rangle$ and
$\sigma_n$ in Fig.~\ref{1} show discontinuities for the same $p_2$
values.

\begin{figure}[b]
\centering \includegraphics[width=0.45\textwidth]{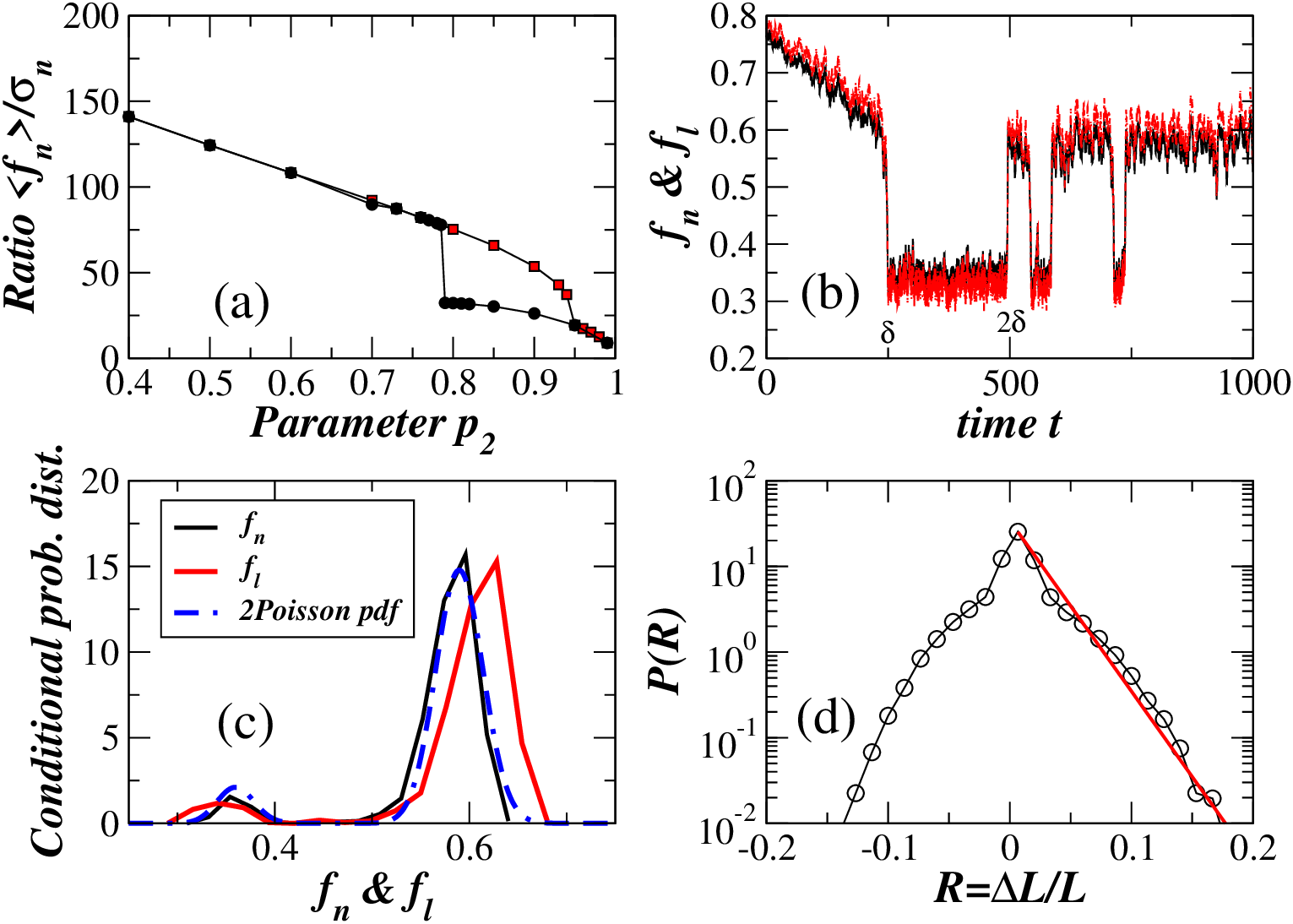}
\caption{Network estimator and forecasting in the (i)-(iii)
  networks. (a) Network estimator, the ratio $\langle f_n \rangle /
  \sigma_n$, exhibits a strong hysteresis---the larger
   $ \langle f_n \rangle/ \sigma_n $, the
  more stable the network. The parameters used are as
  those in Fig. 1(a).  (b) The two fractions, $f_n$ and $f_{\ell}$, 
  of the network   with time-dependent $p_1$ that moves from $I_0$ 
  (Fig.~1(c))
  at time $t=0$ to {\it $X_A$} during the first $\delta = 250$
  steps. From $\delta$ to $2 \delta$, the network stays in {\it $X_A$}.
  Upon reaching {\it $X_A$}, the network phase-flips between mainly
  ``active'', I, and mainly ``inactive'' phases, II.  (c) Moving from $I_0$ to
  {\it $X_A$} after $t=2 \delta$ we calculate two $cdf$s, $C(f_n)$ and
  $C(f_{\ell})$, both exhibiting a highly asymmetric bimodal shape. From
  $C(f_n)$ we can estimate the percentage of (in)active nodes at future
  $t$. Shown also is a combination of two Poissonian distributions, $w_1
  P(\lambda_1) + w_1 P(\lambda_2)$ where  $w_1=0.1$, 
  $w_2=0.9$, $\lambda_1=0.59$ and $\lambda_2=0.36$, where $\lambda_1$ and
  $\lambda_2$ are $\langle f_n \rangle $ values in  I and II.  (d) For
  the number of active links, $ { L(t)} $, we show 
  $P({\Delta L(t)/L}(t))$, and its exponential fit.  }
\label{2}
\end{figure}

We next use the diagnostic power of the (i)-(iii) network  to quantify the
level of its stability. Using the first two moments of
Eq.~(\ref{moment}), we define network risk (volatility) as $\sigma_n 
\equiv \sqrt{ \langle f_n^2 \rangle - (\langle f_n \rangle)^2}$.
Because a network is more stable when $f_n(t)$ is less volatile
($\sigma_n \rightarrow 0$) and when $\langle f_n \rangle$ is as close to
1 as possible, we propose another network stability measure, the
stability network ratio,
\begin{equation} 
\langle f_n \rangle/ \sigma_n,
  \label{ratio}
\end{equation} 
where the larger the ratio, the more stable the
network. Figure~\ref{2}(a) shows that for a (i)--(iii) network the ratio
exhibits hysteresis behavior, e.g., with increasing instability ($p_2
\rightarrow 1$), $\langle f_n \rangle / \sigma_n$ decreases.  When $N$
is large, $\langle f_n \rangle/ \sigma_n  = \sqrt{(1-a)N/a}$ [see
  Eq.~(\ref{moment})]. In practice, if two networks have equal $\langle
f_n \rangle$, but different $\sigma_n$, the one with the larger ratio is
more stable.  Note that a similar first-to-second moment of a  price return
  is proposed in finance to quantify the 
 performance of a financial asset \cite{Sharpe}.  
 Similar signal-to-noise ratio defined as the ratio of mean to standard deviation of a signal is used widely in 
 science and engineering \cite{Bushberg}.

In addition to estimating network volatility, we also need to forecast,
 having the initial configuration of active nodes, 
how many nodes will have failed at any future time $t$.  We allow the
(i)--(iii) network in Fig.~\ref{1}(c), initally at stable state $I_0$,
to move $\delta$ steps (with $p_1$ changing linearly) to a highly
volatile phase-flipping state {\it $X_A$}.  Figure~\ref{2}(b) shows a
representative $f_n(t)$.  We always start from the same initial $I_0$,
perform a large number of simulations [see Fig.~\ref{2}(c)], and obtain
the conditional distribution function (cdf), $C(f_n)$, from which we
calculate the probability ($\int_{0.01q}^1 C(f_n) df_n$) that no more
than $q\%$ nodes will be inactive at $t=2\delta$.  In finance, this
probability approximates the risk that a substantial fraction of
financial system will collapse, the so-called ``systemic risk'' 
\cite{Lo12}.

When we use $f_n$ we are assuming that every node is equally
important. This frequently does not hold for real-world
networks~\cite{AlbertRMP,Dorogotsev,Tak2011}, e.g., when large
banks become dysfunctional they affect the overall financial network
much more than dysfunctioning small banks.  In the (i)--(iii) network
the importance of each node is governed by network topology---the
time-dependent node degree, $k(t)$. 
 A randomly chosen link  is active if both its nodes are active
 and so the  probability that the link is active is $(1 -a )^2$.
 The average number of active links is
\begin{equation}         
    \langle L \rangle  
   =  (1 -a)^2 L_T, 
  \label{loss1}  
\end{equation}    
where  $L_T \equiv 1/2 \Sigma_{i=1}^{N}
k_i$ denotes the total number of links when all links are active. Similar to
Eq.~(\ref{moment}) for a network with $L_T$ links, each with probability
$u \equiv (1-a)^2$ of being active, a $q-$order moment of 
$f_{\ell}(t) = L(t) / L_T$---the fraction of active links---is
\begin{equation} 
 \langle  f^q_{\ell} \rangle 
  \equiv \Sigma_{j=0}^{L_T} (\frac{j}{L_T})^q u^{j} (1-u)^{L_T-j} 
   {L_T \choose j}. 
 \label{momentl}
\end{equation} 
Figure~\ref{2}(b) shows a representative $ f_{\ell}(t)$, and Fig.~\ref{2}(c)
shows $C(f_{\ell})$ (broader than $C(f_n)$) from which we can calculate the
probability ($\int_{0.01q}^1 C(f_{\ell}) d f_{\ell}$) that no more than $q\%$ of
links will be inactive at $t$.  Figure~\ref{2}(d) shows the pdf of the
relative change in $L(t)$ and its exponential fit---which is potentially
important information for network management.  Note that ${\cal L}(t) =
L_T [1 -   f_{\ell}(t)]$ denotes the loss of a network's links.  Using
Eq.~(\ref{loss1}) we obtain $\langle {\cal L} \rangle \equiv L_T -
\langle L \rangle = a(2 -a)L_T $.

\begin{figure}[b]
\centering \includegraphics[width=0.23\textwidth]{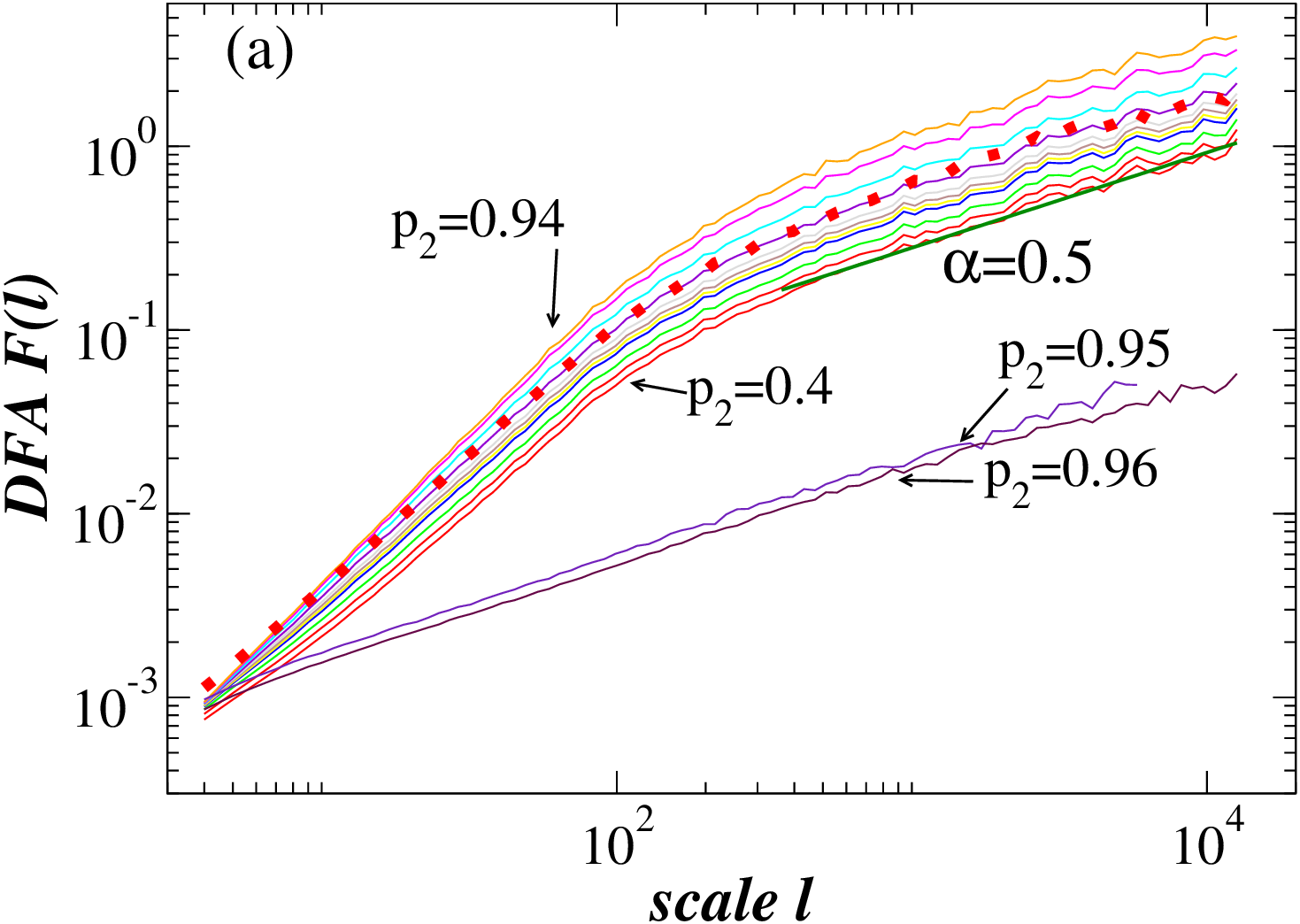}  
\centering \includegraphics[width=0.23\textwidth]{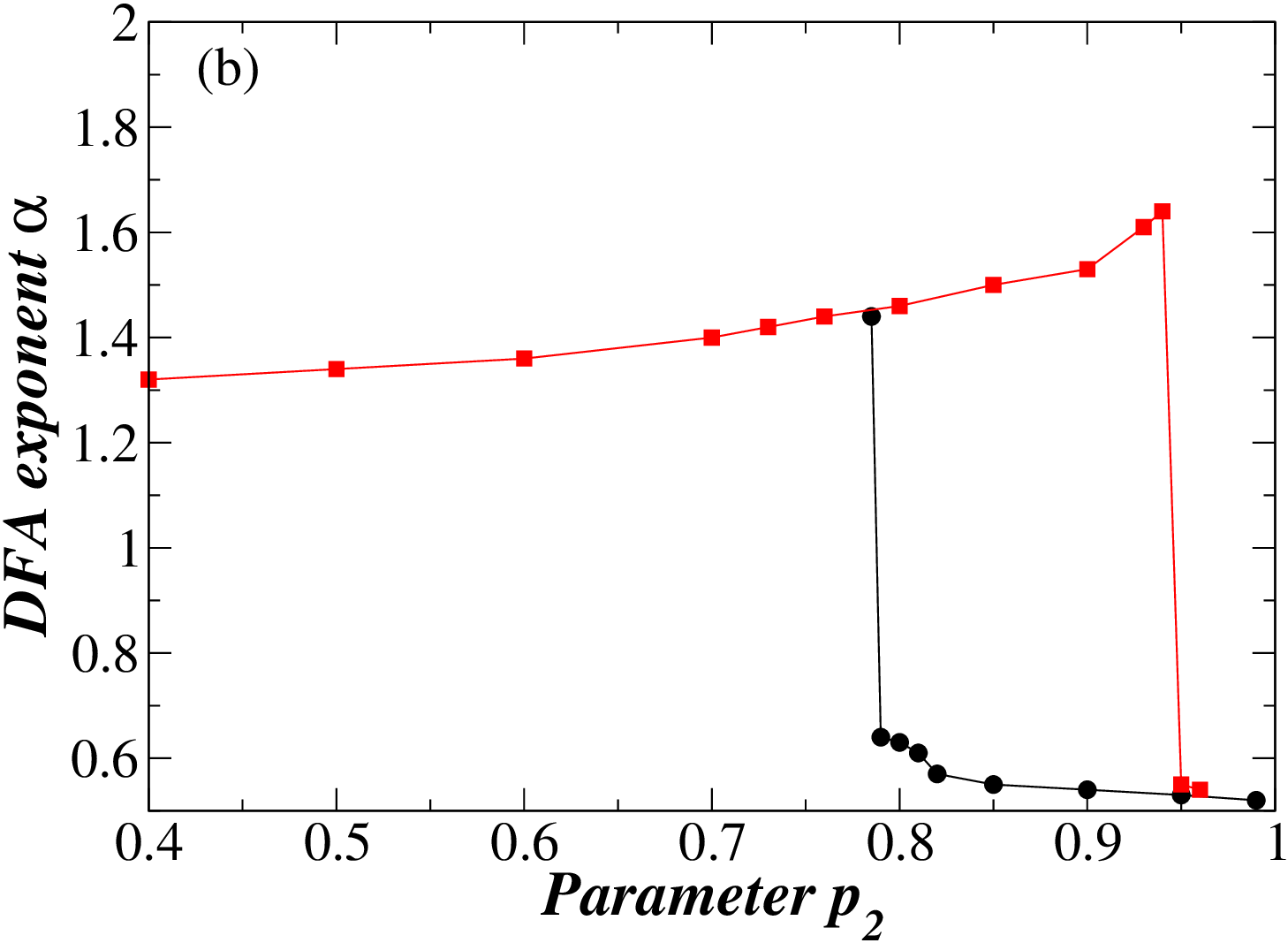}\\
\caption{Emergence of correlations at the network level due to
  failure at the node level. (a) For each time series (constant
  $p_2$) used in
  Fig. 1(a), we show the DFA plot of $f_n(t)$
  vs. scale $\ell$, $F(\ell) \propto (\ell)^{\alpha}$ for the I-II
  transition.  With increasing $p_2$ up to 0.94, $F(\ell)$
  moves upward, accompanied by small (b) increase in the DFA exponent
  $\alpha$---$\alpha$ exhibits a cross-over at a scale 
  $l$ that varies with the recovery times $\tau$ and $\tau'$. Suddenly,
  at $p_2 \approx 0.94$, the DFA exponent drops in a first-order phase
  transition.  We also show correlations  in the fraction
  of externally failed nodes (dotted lines), responsible for correlations in $f_n$. (b)
 Exponent $\alpha$, calculated for both I-II and II-I transitions for
  scales $\ell \le 100$, exhibits a clear hysteresis.  }
\label{3}
\end{figure}
 
The (i)--(iii) network model offers one more potentially important
forecasting property.  Suppose a network set in a state $X_B$ 
(see Fig.~1(c)) within the 
hysteresis regime is predominantly inactive.  Reference~\cite{Antonio03}
defines a local time-dependent parameter $p_{2,\lambda}(t) =
\frac{1}{\lambda} \Sigma_{i=1}^{\lambda}p_{2}(t+1-i)$ as the average
fraction of externally failed nodes over the most recent interval of
length $\lambda$. When $p_{2,\lambda}(t)$ crosses the ``left'' spinodal,
the network shifts from the inactive phase II to the active phase
I. Similarly, $p_{1,\lambda}(t) = \frac{1}{\lambda}
\Sigma_{i=1}^{\lambda}p_{1}(t+1-i)$.  In Ref.~\cite{Antonio03} the pdf
of $p_{2,\lambda}(t)$ ($p_{1,\lambda}(t)$) determines the average
lifetime of the system in I and II.  Here we find that $p_{2}(t)$
follows a binomial distribution that can be approximated for large
samples $n$ with the normal distribution $N(\mu = p_2,
{\sigma}^2=p_2(1-p_2)/n) \equiv P(p_{2}(t))\sim
\exp[-\frac{n(p_{2}(t)-p_2)^2}{2p_2(1-p_2)}]$, where
$n=NE(a(p_1,p_2),k,m)$ [see Eq.~(\ref{a})].  From $p_{2,\lambda}(t) =
\frac{1}{\lambda} \Sigma_{i=1}^{\lambda}p_{2,t+1-i}$ we easily derive
$p_{2,\lambda}(t) =p_{2}(t)/\lambda + p_{2,\lambda}(t-1)-
p_{2}(t-\lambda)/\lambda$.  

Thus, having information about the previous $p_{2,\lambda}$,
$p_{2,\lambda}(t-1)$, we can forecast the current value, where the
closer $p_{2,\lambda}$ is to a spinodal, the larger the probability that
the phase will flip. We quantify this probability using the conditional
distribution function (cdf) $C(p_{2,\lambda}(t))\sim \exp[-\frac{N
    \lambda^2 E(a(p_1,p_2),k,m) (p_{2,\lambda}(t)-p_{2 \lambda}(t-1)+
    p_{2}(t-\lambda)/\lambda - p_2/\lambda)^2}{2p_2(1-p_2)}]$.  
This probability can be used to estimate, given the most recent local
state $p_{2 \lambda}(t-1)$ and $p_{2}(t-\lambda)$, the probability $P(x
\le p_{2s}|p_{2 \lambda}(t-1),p_{2}(t-\lambda))$ that the network will
move from being predominantly inactive, II, to predominantly active,
I---here, as in~\cite{Antonio03}, $p_{2s}$ is a spinodal value where the
network phase-flips from II to I [Fig.~1(c)].  Similarly, if $p_{1s}$
defines a spinodal value at which the network phase-flips from phase I
to II [Fig.~1(c)], from the cdf $C(p_{1 \lambda}(t))\sim
\exp[-\frac{N\lambda^2 (p_{1,\lambda}(t)- p_{1 \lambda}(t-1)+
    p_{1}(t-\lambda)/\lambda -p_1/\lambda)^2}{2p_1(1-p_1)}]$ we can
estimate the probability $P(x \ge p_{1s}|p_{1
  \lambda}(t-1),p_{1}(t-\lambda))$ that the network will fall into the
mainly inactive phase (e.g., as in an economic recession) within
the next period.
 
\begin{figure}[b]

\centering \includegraphics[width=0.23\textwidth]{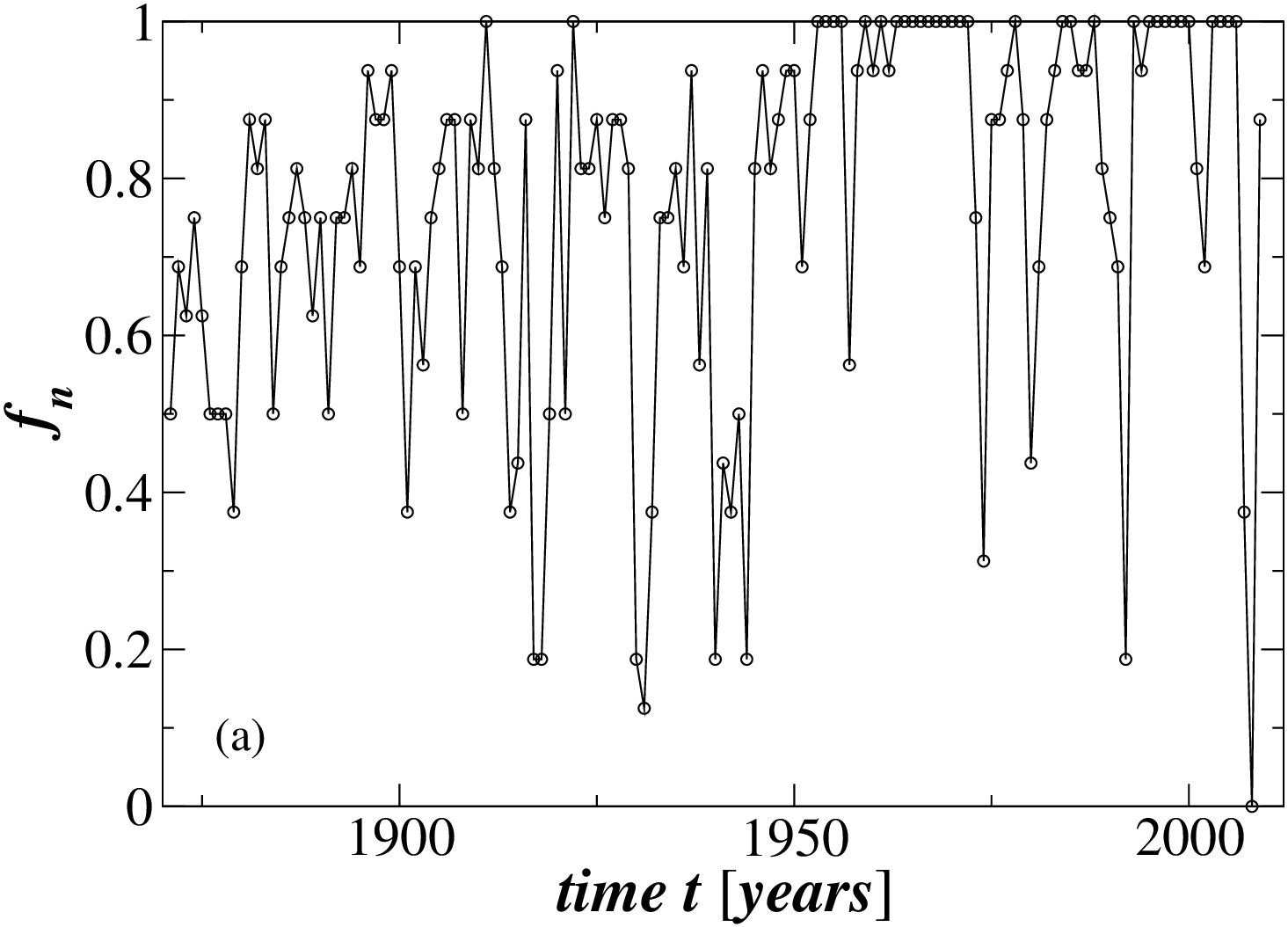}
\centering \includegraphics[width=0.23\textwidth]{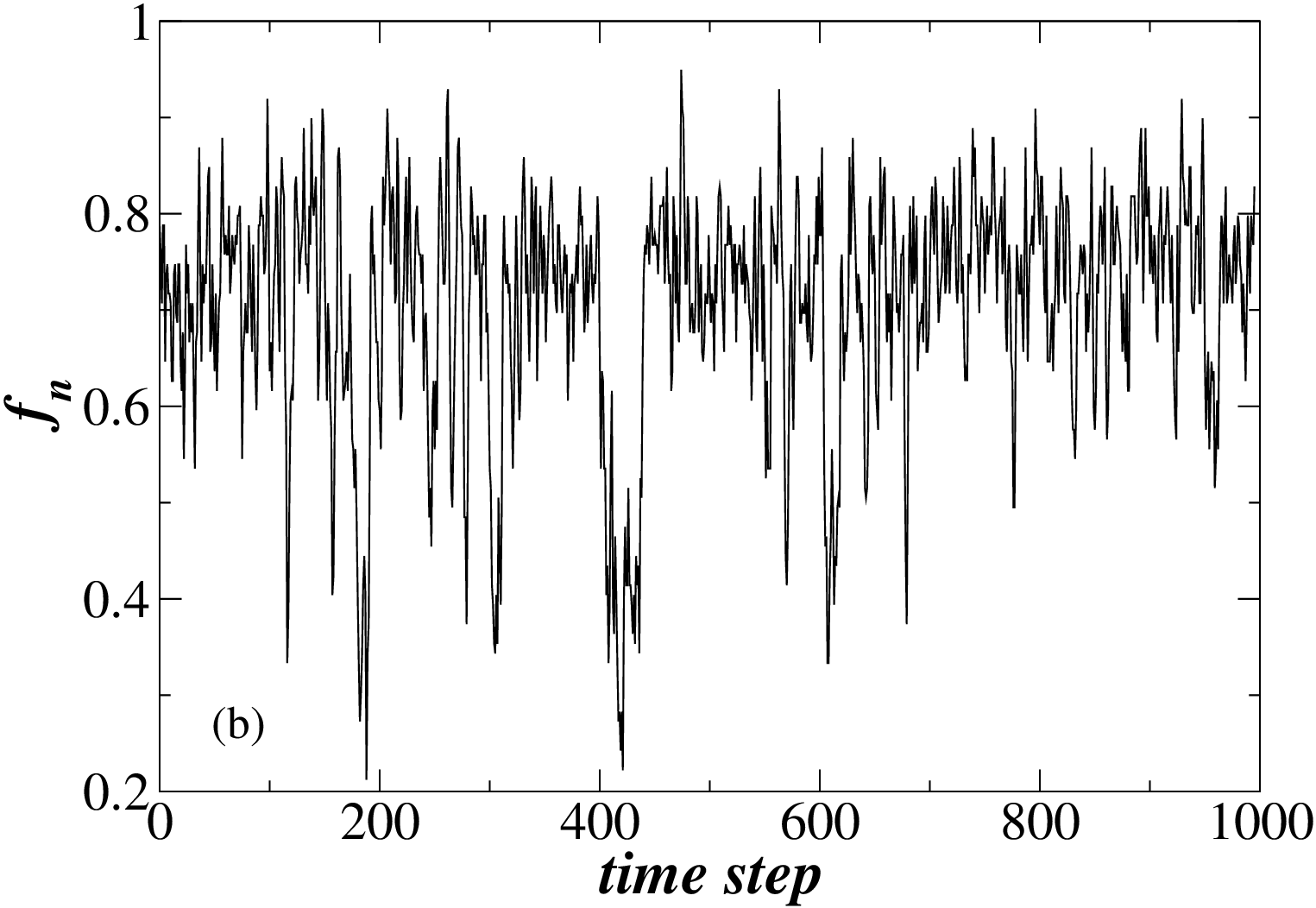}\\
\centering \includegraphics[width=0.22\textwidth]{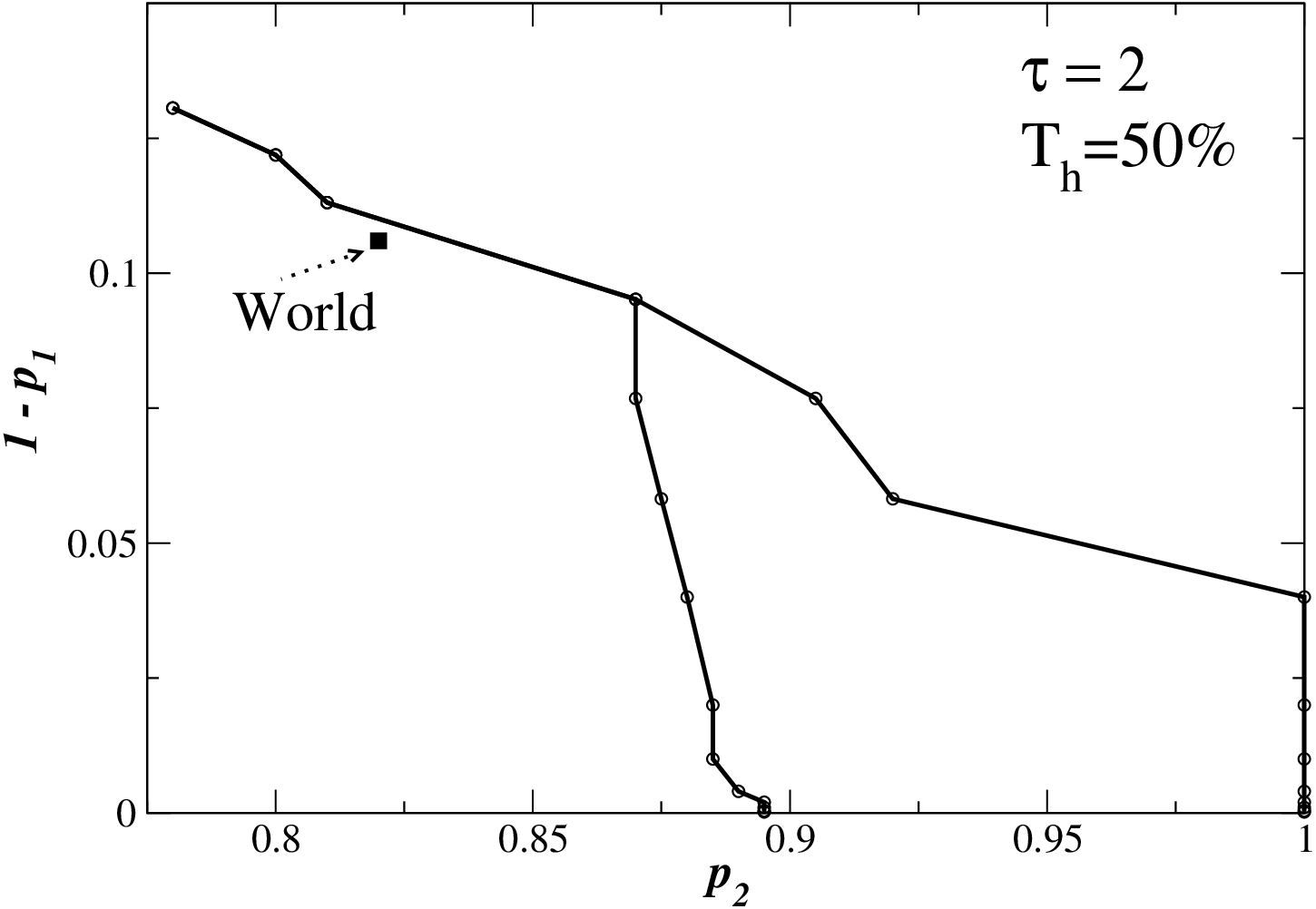}
\centering \includegraphics[width=0.22\textwidth]{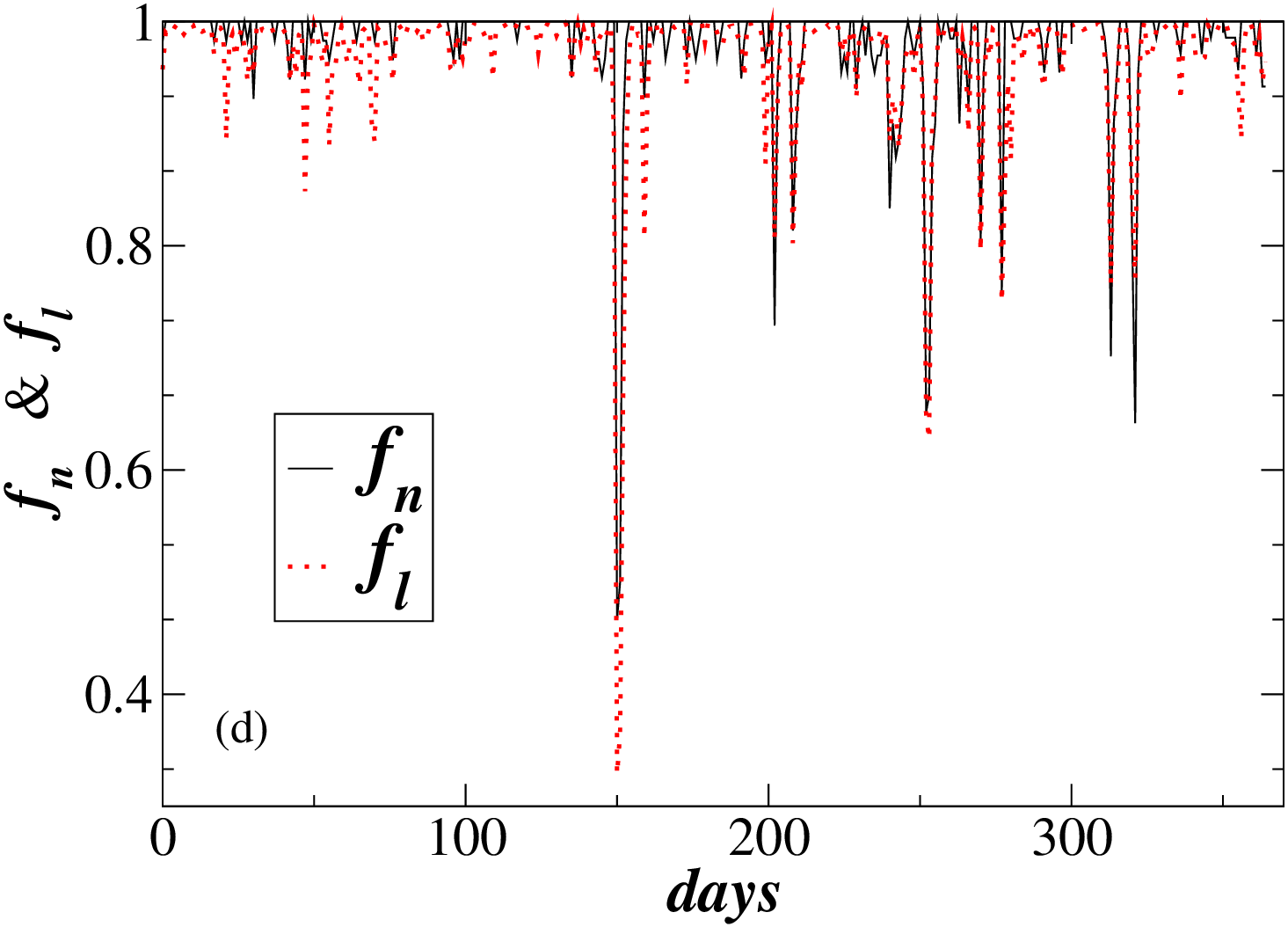}

\caption{Application of dynamical networks.  (a) Economics---how far is
  an economy from a volatile regime?  Fraction of developed countries,
  $f_n(t)$, not in recession (${ \Delta \it gdp} > 0$).  (b) For a
  (i)-(iii) network we show the model's $f_n(t)$ obtained by fitting the
  first and the second moments in (a), where $\tau = 1.3$, $T_h=56~\%$,
  $p_1=0.103$ and $p_2 = 0.77$.  (c) Hysteresis with two spinodals for
  the (i)-(iii) network with $N=100$, $\tau = 2$, and $T_h=50~\%$.  The parameter set is close to the critical line in the
  supercritical region.  (d) Air traffic network.  Fraction of active
  Northeastern US airports, $f_n$, with more than 40\% of canceled
  flights.  We also show $f_{\ell}$.  }
\label{4}
\end{figure}
  
Finally we examine the emerging hysteresis in correlations of $f_n(t)$
due to network dysfunctionality at the node level.  For each time series
$f_n(t)$ ($\tau =~$const) in Figs.~\ref{1}(a) and \ref{1}(b) we apply
detrended fluctuation analysis (DFA) \cite{Peng}---$F^2(l) \propto l^{2
  \alpha}$.  Figure~\ref{3}(a) shows that $f_n(t)$ exhibits finite-range
correlations of the random-walk type ($\alpha \approx 1.5$) with a clear
first-order phase transition, in which a sudden change in the
correlation exponent $\alpha$ occurs when $p_2$ approaches the value at
which we expect network collapse (see Fig.~\ref{1}). An approximate
explanation of the correlations in $f_n(t)$ is that 
%$f_n(t)$ is contributed by only active nodes with $| s_i,S_i \rangle = | 1, 1\rangle $, and thus 
 correlations in $f_n(t)$ and its hysteresis behavior
are due to correlations in the fractions of externally failed nodes
$p_2(t)$ and internally failed nodes $p_1(t)$.  Figure~\ref{3}(a)
confirms this assumption by showing only correlations in $p_2(t)$.  The
existence of hysteresis \cite{Blanchard,Angeli,Das} in Fig.~\ref{3}(b)
indicates that the correlations in collective modes are not the same
when the network approaches network collapse and when the network
recovers---if, e.g., our network models the global economy, then when
the economy moves from ``bad'' to ``good'' years, ``good'' years are
never as good as the previous ``good'' years.

To demonstrate the utility of the (i)--(iii) network model when
analyzing real-world networks, we first analyze a small economic network
of 19 developed countries \cite{datacountry}, and use an output
measurement of trading dynamics, {\it per capita\/} gross domestic
product---{\it gdp}.  For each country and for each year $t$ between
1870 and 2012 \cite{Maddison}, a country (node) is active if the {\it
  gdp} growth is non-negative (if it has been a ``good'' year).
Figure~\ref{4}(a) shows the fraction of active countries $f_n(t)$ that
are becoming increasingly interdependent due to globalization [the
  non-stationary analyzed in Fig.~\ref{2}(b)].  When we disregard this
non-stationarity we find model parameters ($p=0.082\pm0.02$, $p_2= 0.77\pm 0.03$, $\tau = 1.33 \pm 0.5$, and $T_h = 56\pm3\%$)
 for which the $\langle f_n \rangle$ of our model and the
second network moment $\sigma_n$ best fit the empirical moments.
Figure~\ref{4}(b) shows the $f_n(t)$ of the model.  From $p_1 = 1 -
\exp(-p \tau)$ \cite{Antonio03}, $p_1=0.103$ suggests that any randomly
chosen developed country will experience recession (failure)
approximately every ten years, since  $p_1$  represents the average 
fraction of internally failed nodes \cite{Antonio03}. 
  The parameter $p_2 = 0.77$ means that
there is an $\approx 77\%$ probability that a country will undergo
recession if its trading partners have recently experienced recession.
Figure~\ref{4}(c) shows the hysteresis \cite{Blanchard} in $(p_1,p_2)$
space for the (i)--(iii) network model with $\tau = 2$ and 
$T_h = 50\%$. We find that developed countries
%\cite{datacountry} 
with $p_1=0.106\pm0.01$, $p_2= 0.82\pm0.02$ 
 lie close to a critical
hysteresis line--an indication that the world economy is highly
unstable.  We next analyze $f_n(t)$ for 23 Latin American countries, 25 EU
countries, and 25 Asian countries for each year since 1980.  We
calculate the network stability ratio $\langle f_n \rangle/ \sigma_n$ of
Eq.~(\ref{ratio}) for each group and obtain the values $3.25$, $4.15$, and
$6.95$, implying that Asian countries are best performers.
    
We next analyze the airport traffic network \cite{dataair} in the
Northeastern United States (The Library of Congress definition)
comprising 66 airports (nodes), and we consider only those flights
(links) within the Northeast.  For each day during the period 6/1/2012
-- 5/31/2013 we calculate the fraction of failed airports $1 -
f_n(t)$. We arbitrarily define failed airports as those in which more
than $T_h = 40\%$ flights have been canceled for the day.  The air
traffic network in Fig.~\ref{4}(d) shows a much higher level of nodes's
stability  than is typical of economic
networks---rarely does $1 - f_n(t)$ drop to 40\%.  Since air traffic
network is known to be scale-free \cite{Vespignani05}, we apply the 
(i)--(iii)  network model to fit the empirical $\langle f_n
\rangle$ and $\sigma_n$ data to the models's parameters---$p=0.011\pm0.003$ and
$p_2 = 0.92\pm0.03$.  Note that in air traffic network although it is common
for links to fail (for flights to be canceled), airports still function
properly.  This implies that the (i)--(iii) dynamic network model could
be extended to introduce item (iv), the probability, $p_3$, that each
link can fail.

\bigskip

\noindent
BP thanks the support of ZSEM (Grant ZSEM08007). 
 This work is also partially
supported by an NUS Grant "Econophysics and Complex
 Network" (R-144-000-313-133). The BU work is supported by
  NSF (Grant CMMI 1125290).

\end{document}